\documentclass[amsmath,aps,showpacs,a4paper,10pt]{revtex4}

 \usepackage{epsf}
 \usepackage{graphicx}    

\begin{document}

 \newcommand{\be}[1]{\begin{equation}\label{#1}}
 \newcommand{\ee}{\end{equation}}
 \newcommand{\bea}{\begin{eqnarray}}
 \newcommand{\eea}{\end{eqnarray}}
 \def\disp{\displaystyle}

 \def\gsim{ \lower .75ex \hbox{$\sim$} \llap{\raise .27ex \hbox{$>$}} }
 \def\lsim{ \lower .75ex \hbox{$\sim$} \llap{\raise .27ex \hbox{$<$}} }

 \begin{titlepage}

 \begin{flushright}
 arXiv:1505.07546
 \end{flushright}

 \title{\Large \bf Hojman Symmetry in $f(T)$ Theory}

 \author{Hao~Wei\,}
 \email[\,email address:\ ]{haowei@bit.edu.cn}
 \affiliation{School of Physics, Beijing Institute
 of Technology, Beijing 100081, China}

 \author{Ya-Nan~Zhou\,}
 \affiliation{School of Physics, Beijing Institute
 of Technology, Beijing 100081, China}

 \author{Hong-Yu~Li\,}
 \affiliation{School of Physics, Beijing Institute
 of Technology, Beijing 100081, China}

 \author{Xiao-Bo~Zou}
 \affiliation{School of Physics, Beijing Institute
 of Technology, Beijing 100081, China}

 \begin{abstract}\vspace{1cm}
 \centerline{\bf ABSTRACT}\vspace{2mm}
 Today, $f(T)$ theory has been one of the popular modified
 gravity theories to explain the accelerated expansion of the
 universe without invoking dark energy. In this work, we
 consider the so-called Hojman symmetry in $f(T)$ theory.
 Unlike Noether conservation theorem, the symmetry vectors and
 the corresponding conserved quantities in Hojman conservation
 theorem can be obtained by using directly the equations of
 motion, rather than Lagrangian or Hamiltonian. We find that
 Hojman symmetry can exist in $f(T)$ theory, and
 the corresponding exact cosmological solutions are obtained.
 We find that the functional form of $f(T)$ is restricted to
 be the power-law or hypergeometric type, while the universe
 experiences a power-law or hyperbolic expansion. These results
 are different from the ones obtained by using Noether symmetry
 in $f(T)$ theory. Therefore, it is reasonable to find exact
 cosmological solutions via Hojman symmetry.
 \end{abstract}

 \pacs{04.50.Kd, 11.30.-j, 98.80.-k, 95.36.+x}

 \maketitle

 \end{titlepage}

 \renewcommand{\baselinestretch}{1.0}


\section{Introduction}\label{sec1}

As is well known, symmetry plays an important role in the
 theoretical physics. In particular, symmetry is a useful tool
 to select models motivated at a fundamental level, and find
 the exact solutions. One of the well-known symmetries in
 physics is Noether symmetry. In fact, Noether symmetry has
 been extensively used in cosmology and gravity theories,
 for instance, scalar field cosmology~\cite{r1,r2}, $f(R)$
 theory~\cite{r3,r4,r5,r26}, scalar-tensor theory~\cite{r6,r7},
 $f(T)$ theory~\cite{r8,r25}, Gauss-Bonnet gravity~\cite{r9},
 non-minimally coupled cosmology~\cite{r10},
 and others~\cite{r11,r27,r28}. It is worth noting that a
 (point-like) Lagrangian should be given {\it a priori} when
 one uses Noether symmetry in these models.

In the year 1992, Hojman~\cite{r12} proposed a drastically
 new conservation theorem constructed without
 using Lagrangian or Hamiltonian. Unlike Noether conservation
 theorem, the symmetry vectors and the corresponding conserved
 quantities in Hojman conservation theorem can be obtained by
 using directly the equations of motion, rather than Lagrangian
 or Hamiltonian. In general, its conserved quantities and other
 related quantities can be different from the ones in Noether
 conservation theorem.

Here is Hojman conservation theorem~\cite{r12}. We consider
 a set of second order differential equations
 \be{eq1}
 \ddot{q}^{\, i}=F^i\left(q^j,\,\dot{q}^j,\,t\right),~~~~~~~
 i, j=1, \ldots, m
 \ee
 where a dot denotes a derivative with respect to time $t$.
 If $X^i=X^i\left(q^j,\,\dot{q}^j,\,t\right)$ is a symmetry
 vector for Eq.~(\ref{eq1}), it satisfies~\cite{r13,r14}
 \be{eq2}
 \frac{d^2X^i}{dt^2}-\frac{\partial F^i}{\partial q^j}X^j
 -\frac{\partial F^i}{\partial\dot{q}^j}\frac{dX^j}{dt}=0\,,
 \ee
 where
 \be{eq3}
 \frac{d}{dt}=\frac{\partial}{\partial t}+
 \dot{q}^i\frac{\partial}{\partial q^i}+F^i\frac{\partial}
 {\partial\dot{q}^i}\,.
 \ee
 The symmetry vector $X^i$ is defined so that the infinitesimal
 transformation
 \be{eq4}
 \tilde{q}^{\,i}=q^i+\epsilon X^i\left(q^j,\,\dot{q}^j,\,t\right)
 \ee
 maps solutions $q^i$ of Eq.~(\ref{eq1}) into solutions
 $\tilde{q}^{\,i}$ of the same equations (up to $\epsilon^2$
 terms)~\cite{r13,r14}. If the ``force'' $F^i$ satisfies
 (in some coordinate systems)
 \be{eq5}
 \frac{\partial F^i}{\partial\dot{q}^i}=0\,,
 \ee
 then
 \be{eq6}
 Q=\frac{\partial X^i}{\partial q^i}+
 \frac{\partial}{\partial\dot{q}^i}\left(\frac{dX^i}{dt}\right)
 \ee
 is a conserved quantity for Eq.~(\ref{eq1}), namely $dQ/dt=0$.
 In fact, the condition~(\ref{eq5}) can be relaxed~\cite{r12}.
 If the ``force'' $F^i$ satisfies (in some coordinate systems)
 \be{eq7}
 \frac{\partial F^i}{\partial\dot{q}^i}=-\frac{d}{dt}\ln\gamma\,,
 \ee
 where $\gamma=\gamma(q^i)$ is a function of $q^i$, then
 \be{eq8}
 Q=\frac{1}{\gamma}\frac{\partial\left(\gamma X^i\right)}
 {\partial q^i}+\frac{\partial}{\partial\dot{q}^i}
 \left(\frac{dX^i}{dt}\right)
 \ee
 is a conserved quantity for Eq.~(\ref{eq1}). Obviously, if
 $\gamma=const.$, Eqs.~(\ref{eq7}) and (\ref{eq8}) reduce to
 Eqs.~(\ref{eq5}) and (\ref{eq6}). In the proof of this
 conservation theorem~\cite{r12} (see also e.g.~\cite{r24}),
 neither a Lagrangian nor a Hamiltonian is needed, and no
 previous knowledge of a constant of motion for
 system~(\ref{eq1}) is invoked either~\cite{r12}.

In~\cite{r12}, Hojman showed that this conservation theorem
 can drastically restrict the functional form of point
 symmetry transformations, and a harmonic oscillator system
 was considered as a concrete example. Recently, Hojman
 conservation theorem has been used in cosmology and gravity
 theory~\cite{r15,r16}. It is found that Hojman conserved
 quantities exist for a wide range of the potential $V(\phi)$
 of quintessence~\cite{r15} and scalar-tensor
 theory~\cite{r16}, and the corresponding exact cosmological
 solutions have been obtained. As mentioned above, Hojman conserved
 quantities and other related quantities can be different from
 the ones of Noether. In fact, Noether symmetry exists only for
 exponential potential $V(\phi)$~\cite{r1,r6,r7}, while Hojman
 symmetry exists for a wide range of potentials $V(\phi)$,
 including not only exponential but also power-law, hyperbolic,
 logarithmic and other complicated potentials~\cite{r15,r16}.
 Therefore, Hojman symmetry might give rise to new features in
 cosmology and gravity theory.

In the present work, we are interested to consider Hojman
 symmetry in $f(T)$ theory. We try to restrict the functional
 form of $f(T)$, and find the exact cosmological solutions as
 well as the corresponding conserved quantities in $f(T)$
 theory, by using Hojman conservation theorem. In
 Sec.~\ref{sec2}, we briefly review the key points of $f(T)$
 theory at first. In Secs.~\ref{sec3} and \ref{sec4}, we then
 consider Hojman symmetry in $f(T)$ theory with pressureless
 and barotropic matter, respectively. The brief conclusion
 is given in Sec.~\ref{sec5}.


\section{$f(T)$ theory}\label{sec2}

The current accelerated expansion of the universe could be due
 to an unknown energy  component (dark energy) or a modification to
 general relativity (modified gravity)~\cite{r17,r18}. In analogy to
 the well-known $f(R)$ theory, recently $f(T)$ theory~\cite{r19,r20}
 has been proposed as a new modified gravity theory to drive
 the accelerated cosmic expansion without invoking dark energy.
 $f(T)$ theory is a generalized version of the teleparallel gravity
 originally proposed by Einstein~\cite{r21,r22}, in which the
 Weitzenb\"ock connection is used, rather than the Levi-Civita
 connection used in general relativity. Here we briefly review
 the key points of $f(T)$ theory following~\cite{r19,r20}. We
 consider a spatially flat Friedmann-Robertson-Walker (FRW) universe
 whose spacetime is described by
 \be{eq9}
 ds^2=-dt^2+a^2(t)\,d{\bf x}^2\,,
 \ee
 where $a$ is the scale factor. The orthonormal tetrad
 components $e_i(x^\mu)$ relate to the metric through
 \be{eq10}
 g_{\mu\nu}=\eta_{ij}e_\mu^i e_\nu^j\,,
 \ee
 where Latin $i$, $j$ are indices running over 0, 1, 2, 3 for
 the tangent space of the manifold, and Greek $\mu$,~$\nu$
 are the coordinate indices on the manifold, also running over
 0, 1, 2, 3. In $f(T)$ theory, the gravitational action is given by
 \be{eq11}
 {\cal S}_T=\int d^4 x\,|e|\,f(T)\,,
 \ee
 where $|e|={\rm det}\,(e_\mu^i)=\sqrt{-g}\,$, and we use the
 units $16\pi G=\hbar=c=1$ throughout this work. $f(T)$ is a
 function of the torsion scalar $T$, which is defined by
 \be{eq12}
 T\equiv{S_\rho}^{\mu\nu}\,{T^\rho}_{\mu\nu}\,,
 \ee
 with
 \bea
 {T^\rho}_{\mu\nu} &\equiv &-e^\rho_i\left(\partial_\mu e^i_\nu
 -\partial_\nu e^i_\mu\right)\,,\label{eq13}\\
 {K^{\mu\nu}}_\rho &\equiv &-\frac{1}{2}\left({T^{\mu\nu}}_\rho
 -{T^{\nu\mu}}_\rho-{T_\rho}^{\mu\nu}\right)\,,\label{eq14}\\
 {S_\rho}^{\mu\nu} &\equiv &\frac{1}{2}\left({K^{\mu\nu}}_\rho
 +\delta^\mu_\rho {T^{\theta\nu}}_\theta-
 \delta^\nu_\rho {T^{\theta\mu}}_\theta\right)\,.\label{eq15}
 \eea
 For a spatially flat FRW universe, from Eqs.~(\ref{eq12})
 and (\ref{eq9}), we find that
 \be{eq16}
 T=-6H^2\,,
 \ee
 where $H\equiv\dot{a}/a$ is the Hubble parameter, and a
 dot denotes a derivative with respect to cosmic time $t$.
 In $f(T)$ theory, the modified Friedmann
 equation and Raychaudhuri equation read~\cite{r19,r20,r23,r8}
 \bea
 &&12H^2 f_T+f=\rho\,,\label{eq17}\\
 &&48H^2 f_{TT}\dot{H}-f_T\left(12H^2+4\dot{H}\right)-f
 =p\,,\label{eq18}
 \eea
 where a subscript $T$ denotes a derivative with respect
 to $T$, while $\rho$, $p$ are the total energy density and
 pressure, respectively. It is well known that when $f(T)=T$
 the familiar general relativity can be completely recovered.
 Unlike metric $f(R)$ theory whose equations of motion are 4th
 order, the equations of motion (\ref{eq17}) and (\ref{eq18})
 in $f(T)$ theory are 2nd order. This is one of the great
 virtues of $f(T)$ theory. Since Hojman symmetry mentioned
 above mainly deals with the equations of motion rather than
 Lagrangian or Hamiltonian, this virtue makes it easy.


\section{Hojman symmetry in $f(T)$ theory with pressureless matter}\label{sec3}

Here we study Hojman symmetry in $f(T)$ theory. To be simple,
 we first consider a flat FRW universe containing only
 pressureless matter, namely the pressure $p=p_m=0$ and the
 energy density $\rho=\rho_m=\rho_{m0}\,a^{-3}$, where the
 subscript ``0'' indicates the present value of
 the corresponding quantity, and we have set $a_0=1$. Using
 Eqs.~(\ref{eq18}) and (\ref{eq16}), we have
 \be{eq19}
 \dot{H}=\frac{12H^2 f_T+f}{48H^2 f_{TT}-4f_T}\,.
 \ee
 Following~\cite{r15,r16}, we introduce a new variable
 $x\equiv\ln a$, and recast Eq.~(\ref{eq19}) as
 \be{eq20}
 \ddot{x}=-\frac{1}{4}\cdot\frac{f-2Tf_T}{2Tf_{TT}+f_T}
 =F(\dot{x})\,.
 \ee
 Noting that $f$ and its derivatives with respect to $T$ are
 all functions of $T=-6H^2$, the ``force'' $F$ explicitly
 depends only on $\dot{x}=H$. If Hojman symmetry exists in
 $f(T)$ theory, the condition (\ref{eq7}) should be satisfied.
 Noting Eq.~(\ref{eq3}) and $\gamma=\gamma(x)$, we recast
 Eq.~(\ref{eq7}) as
 \be{eq21}
 -\frac{1}{\dot{x}}\frac{\partial F(\dot{x})}{\partial\dot{x}}
 =\frac{\partial}{\partial x}\ln\gamma(x)\,.
 \ee
 Since its left-hand side is a function of $\dot{x}$ only and
 its right-hand side is a function of $x$ only, they must be
 equal to a same constant in order to ensure that
 Eq.~(\ref{eq21}) always holds. For convenience, we let this
 constant be $3/n$ while $n\not=0$, and then Eq.~(\ref{eq21})
 can be separated into two ordinary differential equations
 \be{eq22}
 \frac{\partial}{\partial x}\ln\gamma(x)=\frac{3}{n}\,,~~~~~~~
 \frac{\partial F(\dot{x})}{\partial\dot{x}}=-\frac{3}{n}\,\dot{x}\,.
 \ee
 Thus, we find that
 \bea
 &&\disp\gamma(x)=\gamma_0\,e^{3x/n}\,,\label{eq23}\\[2mm]
 &&\disp F(\dot{x})=-\frac{3}{2n}\,\dot{x}^2+c_0\,,\label{eq24}
 \eea
 where $\gamma_0$ and $c_0$ are both integral constants. In the
 following subsections, we consider the cases of $c_0=0$ and
 $c_0\not=0$, respectively.


\subsection{The case of $c_0=0$}\label{sec3a}

In the case of $c_0=0$, substituting
 $F=-3\dot{x}^2/(2n)=T/(4n)$ into Eq.~(\ref{eq20}), we have
 \be{eq25}
 2T^2 f_{TT}+(1-2n)\,Tf_T+nf=0\,,
 \ee
 which is a differential equation of $f(T)$ with respect to
 $T$. Its solution is given by
 \be{eq26}
 f(T)=c_1 T^n+c_2\sqrt{T}\,,
 \ee
 where $c_1$ and $c_2$ are both integral constants. On the
 other hand, from $\dot{H}=\ddot{x}=F=-3H^2/(2n)$, we get
 \be{eq27}
 H(t)=\frac{2n}{3}(t+c_3)^{-1}\,,
 \ee
 where $c_3$ is an integral constant. Noting that
 $H=\dot{a}/a$, it is easy to obtain
 \be{eq28}
 a(t)=c_4 (t+c_3)^{2n/3}\,,
 \ee
 where $c_4$ is an integral constant. One might set $c_3=0$ by
 requiring $a(t=0)=0$. Note that $n>0$ is required to ensure
 that the universe is expanding. Using Eqs.~(\ref{eq27}) and
 (\ref{eq28}), it is easy to find the dimensionless Hubble parameter
 \be{eq29}
 E\equiv H/H_0=a^{-3/(2n)}=(1+z)^{3/(2n)}\,,
 \ee
 where $z$ is the redshift. Using $\dot{H}=\ddot{x}=F=-3H^2/(2n)$, we
 obtain the deceleration parameter
 \be{eq30}
 q\equiv -\frac{\ddot{a}}{aH^2}=-1-\frac{\dot{H}}{H^2}=
 \frac{3}{2n}-1\,.
 \ee
 When $n>3/2$, the expansion of the universe can
 be accelerated (note that $n<0$ is not acceptable because the
 universe contracts in this case).

Let us turn to the conserved quantity. Following~\cite{r15,r16}, we
 assume that the symmetry vector $X$ does not explicitly depend
 on time. Substituting Eq.~(\ref{eq24}) with $c_0=0$ into
 Eq.~(\ref{eq2}), the equation for $X$ reads
 \be{eq31}
 \frac{\partial^2 X}{\partial x^2}
 -\frac{3\dot{x}}{n}\frac{\partial^2 X}{\partial x\partial\dot{x}}+
 \frac{9\dot{x}^2}{4n^2}\frac{\partial^2 X}{\partial\dot{x}^2}
 +\frac{3}{2n}\frac{\partial X}{\partial x}=0\,.
 \ee
 To solve this equation, we adopt the ansatz
 \be{eq32}
 X=A_0\dot{x}^\alpha e^{\beta x}+A_1\,,
 \ee
 where $A_0$, $A_1$, $\alpha$, $\beta$ are all constants, and
 $\alpha$, $\beta$ cannot be zero at the same
 time. Substituting Eq.~(\ref{eq32}) into Eq.~(\ref{eq31}), we
 find that the solutions have $3\alpha-2n\beta=0$
 or $3\alpha-2n\beta=3$. Substituting Eqs.~(\ref{eq32}) and
 (\ref{eq23}) into Eq.~(\ref{eq8}), the conserved quantity
 $Q$ is given by
 \be{eq33}
 Q=\frac{1}{2n}\left[6A_1-(2+\alpha)
 (3\alpha-2n\beta-3)A_0\dot{x}^\alpha e^{\beta x}\right]\,.
 \ee
 If $3\alpha-2n\beta=3$ or $\alpha=-2$, then $Q=3A_1/n=const.$
 is trivial. If $3\alpha-2n\beta=0$ and $\alpha\not=-2$, we get
 \be{eq34}
 \dot{x}^{2n}e^{3x}=const.
 \ee
 In fact, this conserved quantity can be found in another way.
 Substituting Eq.~(\ref{eq26}) and $T=-6H^2=-6\dot{x}^2$ as
 well as $\rho=\rho_m=\rho_{m0}\,a^{-3}=\rho_{m0}\,e^{-3x}$
 into Eq.~(\ref{eq17}), one can find the same conserved
 quantity given in Eq.~(\ref{eq34}) again. This can be
 regarded as a confirmation of Hojman conservation theorem.


\subsection{The case of $c_0\not=0$}\label{sec3b}

In the case of $c_0\not=0$, substituting Eq.~(\ref{eq24}) into
 Eq.~(\ref{eq20}), and noting $-3\dot{x}^2=T/2$, we obtain
 a differential equation of $f(T)$ with respect to $T$,
 \be{eq35}
 -\frac{1}{4}\cdot\frac{f-2Tf_T}{2Tf_{TT}+f_T}=\frac{T}{4n}+c_0\,.
 \ee
 The corresponding solution for $c_0\not=0$ is given by
 \be{eq36}
 f(T)=c_1(4nc_0)^n\cdot{_2}F_1\left(-\frac{1}{2},\,-n;
 \,\frac{1}{2};\,-\frac{T}{4nc_0}\right)+c_2\sqrt{T}\,,
 \ee
 where ${_2}F_1$ is a hypergeometric function, and $c_1$,
 $c_2$ are both integral constants. In the limit $c_0\to 0$,
 Eq.~(\ref{eq36}) can reduce to Eq.~(\ref{eq26}). Obviously,
 in general $c_0\not=0$ makes difference. From
 $\dot{H}=\ddot{x}=F=-3\dot{x}^2/(2n)+c_0=-3H^2/(2n)+c_0$, we obtain
 \be{eq37}
 H(t)=\sqrt{\frac{2nc_0}{3}}\,\tanh
 \left(\sqrt{\frac{3c_0}{2n}}(t+c_3)\right)\,,
 \ee
 and then
 \be{eq38}
 a(t)=c_4\left[\cosh\left(\sqrt{\frac{3c_0}{2n}}
 (t+c_3)\right)\right]^{2n/3},
 \ee
 where $c_3$ and $c_4$ are both integral constants. Using
 Eqs.~(\ref{eq37}) and (\ref{eq38}), it is easy to obtain the
 Hubble parameter as a function of scale factor
 \be{eq39}
 H(a)=\sqrt{\frac{2nc_0}{3}}\,\tanh\left({\rm arccosh}\left(
 \left(\frac{a}{c_4}\right)^{\frac{3}{2n}}\right)\right)\,,
 \ee
 and then $E=H/H_0$ is ready. On the other hand, using
 $\dot{H}=\ddot{x}=F=-3\dot{x}^2/(2n)+c_0=-3H^2/(2n)+c_0$,
 we also obtain the deceleration parameter
 \be{eq40}
 q\equiv -\frac{\ddot{a}}{aH^2}=-1-\frac{\dot{H}}{H^2}=
 \frac{3}{2n}-1-\frac{c_0}{H^2}\,,
 \ee
 where $H$ is given in Eq.~(\ref{eq37}) or (\ref{eq39}).

Let us turn to the conserved quantity. Substituting Eq.~(\ref{eq24})
 with $c_0\not=0$ into Eq.~(\ref{eq2}), we get the equation for the
 symmetry vector $X$. Some terms with $c_0\not=0$ appear, comparing
 with Eq.~(\ref{eq31}). The ansatz in Eq.~(\ref{eq32}) does
 not work, mainly due to the additional term
 $\sim c_0^2\,\partial^2 X/\partial\dot{x}^2$ (because the order of
 $\dot{x}$ in this term is different from other terms). So, it
 is hard to solve Eq.~(\ref{eq2}) with $c_0\not=0$ to get the
 symmetry vector~$X$. Thus, the task to obtain the conserved
 quantity $Q$ in Eq.~(\ref{eq8}) is also hard. Fortunately,
 there exists another way. Inspired by the discussion below
 Eq.~(\ref{eq34}), we can instead find the
 corresponding conserved quantity by using the modified
 Friedmann equation~(\ref{eq17}). Substituting Eq.~(\ref{eq36}) and
 $T=-6H^2=-6\dot{x}^2$ as well as
 $\rho=\rho_m=\rho_{m0}\,a^{-3}=\rho_{m0}\,e^{-3x}$ into
 Eq.~(\ref{eq17}), it is easy to find the conserved quantity
 \be{eq41}
 \left(\dot{x}^2-\frac{2nc_0}{3}\right)^n e^{3x}=const.
 \ee
 Obviously, it reduces to Eq.~(\ref{eq34}) if $c_0=0$. In fact,
 Eq.~(\ref{eq41}) suggests us to instead adopt another ansatz
 $X=A_0\left(\dot{x}^2+A_3\right)^\alpha e^{\beta x}+A_1$ (or
 something similar) to solve the equation for $X$, and then
 obtain the conserved quantity $Q$ in Eq.~(\ref{eq8}) again.
 Although this is anticipated to work well but the result
 must be the same one given in Eq.~(\ref{eq41}), here we do
 not try the complicated calculation. Let us move forward.


\section{Hojman symmetry in $f(T)$ theory with barotropic matter}\label{sec4}

In this section, we consider a more general case, and assume a
 flat FRW universe containing the so-called barotropic matter,
 i.e. the pressure $p=p_m=w\rho_m$ and the energy density
 $\rho=\rho_m=\rho_{m0}\,a^{-3(1+w)}$, where the constant
 $w\not=-1$ is the equation-of-state parameter (EoS) of
 barotropic matter. Using Eqs.~(\ref{eq18}), (\ref{eq17}) and
 (\ref{eq16}) with $p=w\rho$, we have
 \be{eq42}
 \dot{H}=\frac{(1+w)\left(12H^2 f_T+f\right)}{48H^2 f_{TT}-4f_T}\,,
 \ee
 which can be recast as
 \be{eq43}
 \ddot{x}=-\frac{1+w}{4}\cdot\frac{f-2Tf_T}{2Tf_{TT}+f_T}
 =F(\dot{x})\,.
 \ee
 Noting that $f$ and its derivatives with respect to $T$ are
 all functions of $T=-6H^2$, the ``force'' $F$ explicitly
 depends only on $\dot{x}=H$. If Hojman symmetry exists in
 $f(T)$ theory, the condition (\ref{eq7}) should be satisfied.
 Noting Eq.~(\ref{eq3}) and $\gamma=\gamma(x)$, we recast
 Eq.~(\ref{eq7}) to the one given in Eq.~(\ref{eq21}). Following the
 same derivation between Eqs.~(\ref{eq21})---(\ref{eq24}), we obtain
 the same $\gamma(x)$ and $F(\dot{x})$ given in
 Eqs.~(\ref{eq23}) and (\ref{eq24}). In the
 following subsections, we consider the cases of $c_0=0$
 and $c_0\not=0$, respectively.


\subsection{The case of $c_0=0$}\label{sec4a}

In the case of $c_0=0$, substituting
 $F=-3\dot{x}^2/(2n)=T/(4n)$ into Eq.~(\ref{eq43}), we have
 \be{eq44}
 2T^2 f_{TT}+(1-2\tilde{n})\,Tf_T+\tilde{n}f=0\,,
 \ee
 where $\tilde{n}=n(1+w)$. This is a differential equation
 of $f(T)$ with respect to $T$. Its solution is given by
 \be{eq45}
 f(T)=c_1 T^{\tilde{n}}+c_2\sqrt{T}=c_1 T^{n(1+w)}+c_2\sqrt{T}\,,
 \ee
 where $c_1$ and $c_2$ are both integral constants. On the
 other hand, from $\dot{H}=\ddot{x}=F=-3H^2/(2n)$, we find
 the same $H(t)$, $a(t)$, $E\equiv H/H_0$ and $q$ given
 in Eqs.~(\ref{eq27})---(\ref{eq30}), which have not been
 affected by the EoS of barotropic matter $w$.

Then we turn to the conserved quantity. Following~\cite{r15,r16}, we
 assume that the symmetry vector $X$ does not explicitly depend
 on time. Substituting Eq.~(\ref{eq24}) with $c_0=0$ into
 Eq.~(\ref{eq2}), the equation for $X$ becomes the same one given in
 Eq.~(\ref{eq31}). Following the same derivation between
 Eqs.~(\ref{eq31})---(\ref{eq34}), we obtain the same conserved
 quantity given in Eq.~(\ref{eq34}). At first glance, this
 result is surprising, because it also has not been affected by the
 EoS of barotropic matter $w$. Let us check it in another way.
 Substituting Eq.~(\ref{eq45}) and $T=-6H^2=-6\dot{x}^2$ as well as
 $\rho=\rho_m=\rho_{m0}\,a^{-3(1+w)}=\rho_{m0}\,e^{-3x(1+w)}$
 into Eq.~(\ref{eq17}), one find the same conserved quantity
 given in Eq.~(\ref{eq34}) once again, since the factor
 $1+w=const.$ (note that $w\not=-1$) in both $\rho$ and $f(T)$
 can be removed at the same time. Therefore, it is not so
 surprising that the conserved quantity has the same form given
 in Eq.~(\ref{eq34}).


\subsection{The case of $c_0\not=0$}\label{sec4b}

In the case of $c_0\not=0$, substituting Eq.~(\ref{eq24}) into
 Eq.~(\ref{eq43}), and noting $-3\dot{x}^2=T/2$, we obtain
 a differential equation of $f(T)$ with respect to $T$,
 \be{eq46}
 -\frac{1}{4}\cdot\frac{f-2Tf_T}{2Tf_{TT}+f_T}=
 \frac{T}{4\tilde{n}}+\tilde{c}_0\,,
 \ee
 where $\tilde{n}=n(1+w)$ and $\tilde{c}_0=c_0/(1+w)$. Note
 that $w\not=-1$. The solution for $c_0\not=0$ is given by
 \bea
 f(T)&=&c_1(4\tilde{n}\tilde{c}_0)^{\tilde{n}}
 \cdot{_2}F_1\left(-\frac{1}{2},\,-\tilde{n};
 \,\frac{1}{2};\,-\frac{T}{4\tilde{n}\tilde{c}_0}\right)+
 c_2\sqrt{T}\nonumber\\
 &=&c_1(4nc_0)^{n(1+w)}\cdot{_2}F_1\left(-\frac{1}{2},\,
 -n(1+w);\,\frac{1}{2};\,-\frac{T}{4nc_0}\right)
 +c_2\sqrt{T}\,,\label{eq47}
 \eea
 where ${_2}F_1$ is a hypergeometric function, and $c_1$,
 $c_2$ are both integral constants. On the other hand, from
 $\dot{H}=\ddot{x}=F=-3\dot{x}^2/(2n)+c_0=-3H^2/(2n)+c_0$, we
 obtain the same $H(t)$, $a(t)$, $H(a)$ and $q$ given in
 Eqs.~(\ref{eq37})---(\ref{eq40}), which have not been
 affected by the EoS of barotropic matter $w$.

Let us turn to the conserved quantity. Similar to
 the discussions in the second part of Sec.~\ref{sec3b}, it is
 hard to solve the equation for the symmetry vector $X$,
 mainly due to the additional term
 $\sim c_0^2\,\partial^2 X/\partial\dot{x}^2$. Thus, the task
 to obtain the conserved quantity $Q$ in Eq.~(\ref{eq8}) is
 also hard. Fortunately, there exists another way. Inspired
 by the discussion below Eq.~(\ref{eq34}), we can instead find
 the corresponding conserved quantity by using the modified
 Friedmann equation~(\ref{eq17}). Substituting Eq.~(\ref{eq47})
 and $T=-6H^2=-6\dot{x}^2$ as well as
 $\rho=\rho_m=\rho_{m0}\,a^{-3(1+w)}=\rho_{m0}\,e^{-3x(1+w)}$
 into Eq.~(\ref{eq17}), it is easy to find that the conserved
 quantity has the same form given in Eq.~(\ref{eq41}). Again,
 it also has not been affected by the EoS of barotropic matter $w$.


\section{Conclusion}\label{sec5}
Today, $f(T)$ theory has been one of the popular modified
 gravity theories to explain the accelerated expansion of the
 universe without invoking dark energy. In this work, we
 consider the so-called Hojman symmetry in $f(T)$ theory.
 Unlike Noether conservation theorem, the symmetry vectors and
 the corresponding conserved quantities in Hojman conservation
 theorem can be obtained by using directly the equations of
 motion, rather than Lagrangian or Hamiltonian. We consider
 $f(T)$ theory with pressureless and barotropic matter,
 respectively. We find that Hojman symmetry can exist in $f(T)$
 theory, and the corresponding exact cosmological solutions
 are obtained. The integral constant $c_0$ in the ``force'' $F$
 plays an important role. If $c_0=0$, the corresponding
 functional form of $f(T)$ is restricted to be the power-law
 type, while the universe experiences a power-law expansion.
 The EoS of barotropic matter $w$ does not change these key
 features. It is worth noting that the same results were found
 by using Noether symmetry in $f(T)$ theory~\cite{r8}. However,
 if $c_0\not=0$, the results are drastically changed. In the
 case of $c_0\not=0$, the corresponding functional form of
 $f(T)$ is restricted to be the hypergeometric type, while
 the universe experiences a hyperbolic expansion. The EoS of
 barotropic matter $w$ does not change these key features. Note
 that these new results cannot be found by using Noether
 symmetry in $f(T)$ theory.

As is well known, usually a symmetry leads to a conserved
 quantity. In this work, we also explicitly derived the
 corresponding conserved quantities for Hojman symmetry in
 $f(T)$ theory. They are given in Eqs.~(\ref{eq34}) and
 (\ref{eq41}) for the cases of $c_0=0$ and $c_0\not=0$,
 respectively. Again, the EoS of barotropic matter $w$ does
 not change these key features. Usually, the conserved
 quantities are useful and helpful in many complicated
 cases, since one needs not to deal with the
 middle evolution processes.

In some sense, Hojman symmetry is wider than
 Noether symmetry, and our work confirms this point previously
 found in~\cite{r15,r16} which considered Hojman symmetry in
 quintessence and scalar-tenser theory. So, it is reasonable
 to find exact cosmological solutions via Hojman symmetry. In
 fact, it might open a new window in the field of cosmology and
 gravity theory, and bring new features to them.

Obviously, it is very natural to further consider Hojman
 symmetry in other cosmological models and gravity theories.
 In fact, we found that it is hard to use Hojman symmetry in
 metric $f(R)$ theory, since its equations of motion are 4th
 order. However, the corresponding equations of motion in
 Palatini $f(R)$ theory are 2nd order~\cite{r18} (we thank the
 referee for pointing out this issue), similar to the case of
 $f(T)$ theory. So, it is of interest to consider Hojman
 symmetry in $f(R)$ theory in the Palatini formalism. We
 leave it to the future works.


\section*{ACKNOWLEDGEMENTS}
We thank the anonymous referee for quite useful comments and
 suggestions, which helped us to improve this work. We are
 grateful to Profs. Rong-Gen~Cai and Shuang~Nan~Zhang for
 helpful discussions. We also thank Minzi~Feng, as well as
 Zu-Cheng~Chen, Jing~Liu and Xiao-Peng~Yan for kind help
 and discussions. This work was supported in part by NSFC
 under Grants No.~11175016 and No.~10905005.

\renewcommand{\baselinestretch}{1.1}


\end{document}